\documentclass[12pt]{article}
\usepackage{amsmath,amssymb}

\newcommand{\be}{\begin{equation}}
\newcommand{\ee}{\end{equation}}
\newcommand{\bea}{\begin{array}}
\newcommand{\ea}{\end{array}}
\newcommand{\beqa}{\begin{eqnarray}}
\newcommand{\eeqa}{\end{eqnarray}}
\newcommand{\bean}{\begin{eqnarray*}}
\newcommand{\eean}{\end{eqnarray*}}

\newcommand{\nn}{\nonumber}
\def\sqr#1#2{{\vcenter{\vbox{\hrule height.#2pt
        \hbox{\vrule width.#2pt height#1pt \kern#1pt
          \vrule width.#2pt}
        \hrule height.#2pt}}}}

\def\half{\frac{1}{2}}
\def\tr{\mathop{\rm Tr}\nolimits}
\def\BI{{\rm 1\!l}}
\def\CP{{\mathbb C}P}

\newcommand{\gapproxeq}{\lower .7ex\hbox{$\;\stackrel{\textstyle
>}{\sim}\;$}}
\newcommand{\lapproxeq}{\lower .7ex\hbox{$\;\stackrel{\textstyle
<}{\sim}\;$}}

\newcounter{appendice}

\def\thebibliography#1{{\bf REFERENCES\markboth
 {REFERENCES}{REFERENCES}}\list
 {[\arabic{enumi}]}{\settowidth\labelwidth{[#1]}\leftmargin\labelwidth
 \advance\leftmargin\labelsep
 \usecounter{enumi}}
 \def\newblock{\hskip .11em plus .33em minus -.07em}
 \sloppy
 \sfcode`\.=1000\relax}

\begin{document}
\begin{titlepage}
\title{{\small\hfill SU-4252-763, DFUP-02-13}\\
Fuzzy Nambu-Goldstone Physics  }
\author{
A. P. Balachandran$^*$ and
G. Immirzi$^\dagger$\\
~\\
{\small\it $^*$ Physics Department,
Syracuse University}\\
{\small\it Syracuse NY 13244-1130, USA}\\
~\\
{\small\it $^\dagger$ Dipartimento di Fisica, Universit\`a di Perugia
{\small\rm and} INFN, Sezione di Perugia,}\\
{\small\it Perugia, Italy}\\
}

\maketitle

\begin{abstract}
In spacetime dimensions larger than 2,
whenever a global symmetry $G$ is spontaneously broken to
a subgroup $H$, and $G$ and $H$ are Lie groups, there
are Nambu -Goldstone modes described by fields with values in $G/H$.
In two-dimensional spacetimes as well, models where fields take values
in $G/H$ are of considerable interest even though in that case there is
no spontaneous breaking of continuous symmetries.
We consider such models when the world sheet is a two-sphere and describe
their fuzzy analogues for $G=SU(N+1),\; H=S(U(N-1)\otimes U(1))\simeq U(N)$
and $G/H=\CP^N$. More generally our methods give fuzzy versions of
continuum models on $S^2$ when the target spaces are Grassmannians
and flag manifolds described by $(N+1)\times(N+1)$ projectors of rank
$\le (N+1)/2$.
These fuzzy models are finite-dimensional matrix models
which nevertheless retain all the essential continuum topological features
like solitonic sectors. They seem well-suited for numerical work.
\end{abstract}
\end{titlepage}
\section{Introduction}

In spacetime dimensions larger than 2,
whenever a global symmetry $G$ is spontaneously broken to a subgroup $H$,
and $G$ and $H$ are Lie groups, there are
massless Nambu-Goldstone modes with values in the coset space $G/H$.
Being massless,
they dominate low energy physics as is the case with pions in strong
interactions and
phonons in crystals. Their theoretical description contains new concepts
because $G/H$ is not a vector space.

Such $G/H$ models have been studied extensively in 2-d physics,
even though in that case there is no spontaneous breaking of
continuous symmetries.
A reason is that they are often tractable nonperturbatively in the
two-dimensional
context, and so can be used to test ideas suspected to be true in
higher dimensions.
A certain amount of numerical work has also been done on such 2-d
models to control
conjectures and develop ideas, their discrete versions having been
formulated for this purpose.

Our work in this paper is on new discrete approximations to $G/H$ models.
We focus on
two-dimensional Euclidean quantum field theories with target space
$G/H=SU(N+1)/U(N)={\mathbb C}P^N$. The novelty in our approach is that
our discretizations are based on
fuzzy physics \cite{madore}
and noncommutative geometry \cite{connes}. Fuzzy physics has striking
elegance because
it preserves the symmetries of the continuum and because techniques of
noncommutative
geometry give us powerful tools to describe continuum topological
features. But
its numerical efficiency has not been tested \cite{sachin}.
We got into this investigation with this mind, our idea being to write fuzzy
$G/H$ models in a form adapted to numerical work.

This is not the first paper on fuzzy $G/H$. In \cite{baez}, a particular
description
based on projectors and their orbits was discretized. We shall refine
that work considerably in this paper. Also in the continuum there
is another way to
approach $G/H$, namely as gauge theories with gauge invariance under $H$ and
global symmetry
under $G$ \cite{baletal}. This approach is extended here to fuzzy
physics. Such a fuzzy gauge theory involves the decomposition of
projectors in terms of partial isometries \cite{wegge} and brings
new ideas into this
field. It is also very pretty. It is equivalent to the projector method
as we shall also see.

Parallel work on fuzzy $G/H$ model and their solitons is being completed
by  Govindarajan  and Harikumar \cite{harikumar}. A different
treatment, based on the Holstein-Primakoff realization of the $SU(2)$
algebra, has been given in \cite{chan}. A more general approach to
these models on non-commutative spaces was proposed in \cite{dabrowski}.

The first two sections describe the standard $\CP^1$-models on $S^2$.
In section 2 we discuss it using projectors, while in section 3 we
reformulate the discussion in such a manner that transition to fuzzy
spaces is simple.
Sections 4 and 5 adapt the previous sections to fuzzy spaces.

Long ago, general $G/H$-models on $S^2$ were written as gauge theories
\cite{baletal}. Unfortunately their fuzzification for generic $G$ and
$H$ eludes us.
Generalization of the considerations here to the case where $S^2\simeq\CP^1$
is replaced with $\CP^N$, or more generally Grassmannians and flag manifolds
associated with $(N+1)\times(N+1)$ projectors of rank $\le(N+1)/2$,
 is easy as we briefly show in the concluding section 6.
But extension to higher ranks remains a problem.

\section{$CP^1$ models and Projectors}
\setcounter{equation}{0}
Let the unit vector $x=(x_1,x_2,x_3)\in {\mathbb R}^3$ describe a point
of $S^2$. The field $n(x)$ in the ${\mathbb C}P^1$-model is a map from
$S^2$ to $S^2$:
\be
 n=(n_1,n_2,n_3)\,:\ x\rightarrow\ n(x)\;\in\,{\mathbb R}^3,\quad
 n(x)\cdot n(x):=\sum_a n_a(x)^2=1\;.
 \label{ngi}
\ee
These maps $n$ are classified by their winding number ${\kappa}\in{\mathbb Z}$:
\be
 \kappa=\frac{1}{8\pi}\int_{S^2}\epsilon_{abc}\,n_a(x)\,dn_b(x)\,dn_c(x)\;.
 \label{ngii}
\ee
 That $\kappa$ is the winding of the map can be seen taking spherical
 coordinates $(\Theta,\Phi)$ on the target sphere $ ( n^2=1 ) $ and using the
 identity $\sin\Theta d\Theta\,d\Phi=\half\epsilon_{abc}n_a dn_b\,dn_c$. 
We omit wedge symbols in forms.

We can think of $n$ as the field at a fixed time $t$ on a (2+1)-dimensional
manifold where the spatial slice is $S^2$. In that case, it can describe a
field of spins, and the fields with $\kappa\ne 0$ describe solitonic
sectors.
We can also think of it as a field on Euclidean spacetime $S^2$. In that
case, the fields with $\kappa\ne 0$ describe instantonic sectors.

Let $\tau_a$ be the Pauli matrices. Then each $n(x)$ is associated with the
projector
\be
  P(x)=\half (1+\vec\tau\cdot\vec n(x))\;.
 \label{ngiii}
\ee
Conversely, given a $2\times 2$ projector $P(x)$ of rank 1, we can write
\be
 P(x)=\half (\alpha_0(x)+\vec \tau\cdot\vec \alpha(x))\;.
 \label{ngiv}
\ee
Using $\tr P(x)=1,\ P(x)^2=P(x)$ and $P(x)^\dagger =P(x)$, we get
\be
 \alpha_0(x)=1,\quad \vec\alpha(x)\cdot\vec\alpha(x)=1,
 \quad \alpha^*_a(x)=\alpha_a(x)\;.
 \label{ngv}
\ee
Thus $\CP^1$-fields on $S^2$ can be described either by $P$ or
by $ n_a=\tr (\tau_a\,P)$ \cite{wojciech}.

In terms of $P$, $\kappa$ is
\be
 \kappa=\frac{1}{2\pi i}\int_{S^2}\tr P\,(dP)\,(dP)\;.
 \label{ngvi}
\ee

There is a family of projectors, called Bott projectors \cite{mignaco,landi}
which play a central role in our approach. Let
\be
 z=(z_1,z_2),\quad |z|^2:= |z_1|^2+|z_2|^2=1\; .
 \label{ngvii}
\ee
The $z$'s are points on $S^3$. We can write $x\in S^2$ in terms of $z$:
\be
 x_i(z)=z^\dagger\tau_i z
 \label{ngviii}
\ee
The Bott projectors are
\beqa
  P_\kappa(x)=v_\kappa(x)v_\kappa^\dagger(z),\quad
  &v_\kappa(z)&=\left[\begin{matrix} z_1^\kappa\\z_2^\kappa \\ 
\end{matrix}\right]\frac{1}{\sqrt{Z_\kappa}}\quad 
\hbox{if}\ \kappa\ge 0\ ,\nn\\
&  Z_k&\equiv|z_1|^{2|\kappa|}+|z_2|^{2|\kappa|}\ ,\nn\\
 &v_\kappa(z)&=
 \left[\begin{matrix}z_1^{*|\kappa|}\\z_2^{*|\kappa|} \\ \end{matrix}\right]
\frac{1}{\sqrt{Z_\kappa}}
\quad  \hbox{if}\ 
\kappa<0 \ .
 \label{ngix}
\eeqa
The field $n^{(\kappa)}$ associated with $P_\kappa$ is given by
\be
 n^{(\kappa)}_a(x)=\tr\tau_a P_\kappa(x)=v_\kappa^\dagger(z)\tau_a v_\kappa(z)
 \ .\label{ngx}
\ee
Under the phase change $z\rightarrow z e^{i\theta}$, $v_\kappa(z)$ changes
$v_\kappa(z)\rightarrow v_\kappa(z)e^{i\kappa\theta}$, whereas $x$ is 
invariant.
As this phase cancels in $v_\kappa(z)v_\kappa^\dagger(z)$, $P_\kappa$ is a 
function of
$x$ as written.

The $\kappa$ that appears in eqs.(\ref{ngix})(\ref{ngx}) is the winding 
number as the explicit calculation of section 3 will show. 
But there is also the following argument.

In the map $z\rightarrow v_\kappa(z)$, for $\kappa=0$, all of $S^3$ and 
$S^2$ get mapped to a point, giving zero winding number. 
So, consider $\kappa>0$. Then the points
\be
 \left( z_1e^{i\frac{2\pi}{\kappa}(l+m)},\, z_2e^{i\frac{2\pi}{\kappa}m} 
\right) 
,\quad l,m\in\;\{ 0,1,..,\kappa-1\}
\nn
\ee
have the same image. But the overall phase $e^{i\frac{2\pi}{\kappa}m}$ of
$z$ cancels out in $x$. Thus, generically $\kappa$ points of $S^2$
(labeled by $l$) have the same projector $P_\kappa(x)$, giving winding
number $\kappa$. As for $\kappa<0$, we get $|\kappa|$ points of $S^2$ 
mapped to the same $P_\kappa(x)$. But because of the complex conjugation in 
eq.(\ref{ngix}),
there is an orientation-reversal in map giving $-|\kappa|=\kappa$ as 
winding numbers. One way to see this is to use 
\be
 P_{-|\kappa|}(x)=P_{|\kappa|}(x)^T
 \label{ngxi}
\ee
Substituting this in (\ref{ngvi}), we can see that $P_{\pm|\kappa|}$
have opposite winding numbers.

The general projector ${\cal P}_\kappa(x)$ is the gauge transform
of $P_\kappa(x)$:
\be
 {\cal P}_\kappa(x)=U(x)P_\kappa(x)U(x)^\dagger
 \label{ngxii}
\ee
where $U(x)$ is a unitary $2\times 2$ matrix. Its $n^{(\kappa)}$ is also
given by (\ref{ngx}), with $P_\kappa$ replaced by ${\cal P}_\kappa$.
 The winding number is
unaffected by the gauge transformation. That is because $U$ is a map from
$S^2$ to $U(2)$ and all such maps can be deformed to identity since
$\pi_2(U(2))=\{\hbox{identity}\ e\}$.

The identity
\be
 {\cal P}_\kappa(d{\cal P}_\kappa)=(d{\cal P}_\kappa)(\BI-{\cal P}_\kappa)
 \label{ngxiii}
\ee
which follows from ${\cal P}_\kappa^2={\cal P}_\kappa$, is valuable when
working with projectors.


\section{ An Action}
\setcounter{equation}{0}

Let ${\cal L}_i=-i(x\wedge\nabla)_i$ be the angular momentum operator.
Then a Euclidean action in the $\kappa$-th topological sector
 for $n^{(\kappa)}(x)$ (or a static Hamiltonian in the (2+1)
 picture) is
\be 
 S_\kappa=-\frac{c}{2}\int_{S^2}d\Omega\, ({\cal L}_i n^{(\kappa)}_b)({\cal 
L}_i n^{(\kappa)}_b)
\ ,\quad c=\ \hbox{a positive constant,}
\label{ngxiv}
\ee
where $d\Omega$ is the $S^2$ volume form $d\cos\theta\, d\varphi$. We can also
write
\be 
 S_\kappa=-c\int_{S^2}d\Omega\, \tr\,({\cal L}_i{\cal P}_\kappa)({\cal 
L}_i{\cal P}_\kappa)
\; .
\label{ngxv}        
\ee
The following identities, based on (\ref{ngxiii}), are also useful:
\be
\tr\, {\cal P}_\kappa({\cal L}_i{\cal P}_\kappa)^2=
\tr\,({\cal L}_i{\cal P}_\kappa)
(\BI-{\cal P}_\kappa)({\cal L}_i{\cal P}_\kappa)=\tr
(\BI-{\cal P}_\kappa)({\cal L}_i{\cal P}_\kappa)^2=\half \tr({\cal L}_i{\cal 
P}_\kappa)^2
\label{ngxvb}
\ee
Hence
\be
S_\kappa=-2c\int_{S^2} d\Omega\,\tr{\cal P}_\kappa\; {\cal L}_i{\cal 
P}_\kappa\;{\cal L}_i{\cal P}_\kappa
\label{ngxvi}
\ee

The Euclidean functional integral for the actions $S_\kappa$ is
\be
 Z(\psi)=\sum_\kappa e^{i\kappa\psi}\int{\cal D\,P}_\kappa e^{-S_\kappa}
 \label{ngxviii}
\ee
where the angle $\psi$ is induced by the instanton sectors as in QCD.

Using the identity $dP=-\epsilon_{ijk}\,dx_i\,x_j\,i{\cal L}_k P$, we can
rewrite the definiton (\ref{ngii}) or (\ref{ngvi}) of the winding number as
\beqa
\kappa&=&\frac{1}{8\pi}\int_{S^2}d\Omega\, \epsilon_{ijk}x_i\,\epsilon_{abc}
n_a^{(\kappa)}\,i{\cal L}_jn_b^{(\kappa)}\,  i{\cal L}_kn_c^{(\kappa)}=
\label{ngxviiib}\\      
 &=& \frac{1}{2\pi i}\int_{S^2}d\Omega\tr {\cal P}_\kappa\,
\epsilon_{ijk}\,x_i\,i{\cal L}_j{\cal P}_\kappa\,i{\cal L}_k{\cal P}_\kappa\ .
\label{ngxviiia}  
\eeqa
The Belavin-Polyakov bound \cite{belavin} 
\be
S_\kappa\ge 4\pi\, c\,|\kappa|
\label{ngxvib}
\ee
 follows from (\ref{ngxviiib}) on integration of
\be
(i{\cal L}_i 
n_a^{(\kappa)}\pm\epsilon_{ijk}x_j\,\epsilon_{abc}\,n_b^{(\kappa)}\, 
i{\cal L}_k n_c^{(\kappa)})^2\ge 0\ ,
\label{nxvii}
\ee
or from (\ref{ngxviiia})  on integration of
\be
\tr\big({\cal P}_\kappa(i{\cal L}_i{\cal P}_\kappa)\pm i\epsilon_{ijk}\,x_j
{\cal P}_\kappa(i{\cal L}_k{\cal P}_\kappa)\big)^\dagger
\big({\cal P}_\kappa(i{\cal L}_i{\cal P}_\kappa)\pm i\epsilon_{ij'k'}\,x_{j'}
{\cal P}_\kappa(i{\cal L}_{k'}{\cal P}_\kappa)\big)
\ge 0\ .
\label{ngxviiic}
\ee
From this last form it is easy to rederive the bound in a way better adapted 
to fuzzification. Using Pauli matrices $\{\sigma_i\}$ 
 we first rewrite (\ref{ngxvi}) and (\ref{ngxviiia}) as
\beqa
S_\kappa&=&c\int_{S^2}d\Omega\tr\,{\cal P}_\kappa 
(i\sigma\cdot {\cal L}\,{\cal P}_\kappa) 
(i\sigma\cdot {\cal L}\,{\cal P}_\kappa)\ ,\nn\\ 
\kappa&=&\frac{-1}{4\pi}\int_{S^2}d\Omega\tr\big(\sigma\cdot x\, {\cal P}_k
(i\sigma\cdot{\cal L}\,{\cal P}_k)(i\sigma\cdot{\cal L}\,{\cal
P}_k)\big)\ .
\label{nglxviiic}
\eeqa
The trace is now over ${\mathbb C}^2\times{\mathbb C}^2={\mathbb C}^4$,
where $\tau_a$ acts on the first ${\mathbb C}^2$ and $\sigma_i$ on
the second ${\mathbb C}^2$ (so they are really
$\tau_a\otimes\BI$ and $\BI\otimes\sigma_i$)
Then, with $\epsilon_1,\epsilon_2=\pm 1$,
\be
\frac{1+\epsilon_2 \tau\cdot n^{(\kappa)}}{2}\sigma_i\big(
(i{\cal L}_i{\cal P}_\kappa)+\epsilon_1 
i\epsilon_{ijk}\,x_j
(i{\cal L}_k{\cal P}_\kappa)\big)
=(1+\epsilon_1\sigma\cdot x)\frac{1+\epsilon_2 \tau\cdot n^{(\kappa)}}{2}
(i\sigma\cdot{\cal L}{\cal P}_\kappa)\ ,
\label{nglxviiid}
\ee
since $x\cdot{\cal L}=0$. The inequality (\ref{ngxviiic}) is equivalent to
\be
\tr\,\left[\frac{1+\epsilon_1\sigma\cdot x}{2}\,
\frac{1+\epsilon_2 \tau\cdot n^{(\kappa)}}{2}\,
(i\sigma\cdot{\cal L}{\cal P}_\kappa)\right]^\dagger
\left[\frac{1+\epsilon_1\sigma\cdot x}{2}\,
\frac{1+\epsilon_2 \tau\cdot n^{(\kappa)}}{2}\,
(i\sigma\cdot {\cal L}{\cal P}_\kappa)\right]   \ge 0\ ,
\label{nglxviiie}
\ee
from which (\ref{ngxvib}) follows by integration.

\section{$\CP^1$-models and Partial Isometries}
\setcounter{equation}{0}

If ${\cal P}(x)$ is a rank 1 projector at each $x$, we can find its 
normalized eigenvector $u(z)$:
\be
  {\cal P}(x)u(z)=u(z)\, ,\quad u^\dagger(z)u(z)=1\; .
 \label{ngxix}
\ee
Then
\be
{\cal P}(x)=u(z)u^\dagger(z)\; .
 \label{ngxx}
\ee
If ${\cal P}={\cal P}_\kappa$ an example of $u$ is $v_\kappa$. 
$u$ can be a function of $z$, changing by
a phase under $z\rightarrow z e^{i\theta}$. Still, ${\cal P}$ will depend
only on $x$.

We can regard $u(z)^\dagger $ (or a slight generalization of it) as an
example of a partial isometry \cite{wegge} in the algebra ${\cal A}=
C^\infty(S^3)\otimes_{\mathbb C}Mat_{2\times 2}({\mathbb C})$
of $2\times 2$ matrices with coefficients in $C^\infty(S^3)$.
A partial isometry in a $*-$algebra $A$ is an element 
${\cal U}^\dagger\in A$
such that ${\cal U\,U}^\dagger$ is a projector; ${\cal U\,U}^\dagger$ 
is the {\it support projector} of ${\cal U}^\dagger$. 
It is an isometry if ${\cal U}^\dagger\,{\cal U}=\BI$. With
\be
 {\cal U}=\left(\begin{matrix} u_1&0\\u_2&0 \end{matrix}\right) 
\in{\cal A} ,
 \label{ngxxi}
\ee
we have 
\be
 {\cal P}={\cal U\,U}^\dagger
 \label{ngxxii}
\ee
so that  ${\cal U}^\dagger$ is a partial isometry.

We will be free with language and  also call $u^\dagger$ as a 
partial isometry.

The partial isometry for $P_\kappa$ is $v^\dagger_\kappa$.

Now consider the one-form
\be
 A_\kappa=v^\dagger_\kappa\, dv_\kappa\;.
 \label{ngxxiii}
\ee
Under $z_i\rightarrow z_ie^{i\theta(x)}$, $A_\kappa$ transforms like a
connection:
\be
A_\kappa\rightarrow A_\kappa+i\kappa\,d\theta
\nn\ee
($A_\kappa$ are connections for $U(1)$ bundles on $S^2$ for Chern numbers
$\kappa$, see later.) Therefore
\be
 D_\kappa=d+A_\kappa
 \label{ngxxiv}
\ee
is a covariant differential, transforming under $z\rightarrow 
z e^{i\theta}$ as
\be
 D_\kappa\rightarrow e^{i\kappa\theta}D_\kappa  e^{-i\kappa\theta}
 \label{ngxxv}
\ee
and
\be
 D_\kappa^2=dA_\kappa
 \label{ngxxvi}
\ee
is its curvature.

At each $z$, there is a unit vector $w_\kappa(z)$ perpendicular to
$v_\kappa(z)$.
An explicit realization of $w_\kappa(z)$ is given by
\be
w_{\kappa,\alpha}=i\tau_{2\,\alpha\beta}\,v^*_{\kappa,\beta}:=
\epsilon_{\alpha\beta}\,v^*_{\kappa,\beta}
 \label{ngxxvii}
\ee
Since $w^\dagger_\kappa v_\kappa=0$,
\be
 B_\kappa=w^\dagger_\kappa\,dv_\kappa\ ,\quad 
B_\kappa^*=(dv^\dagger_\kappa)w_\kappa=-v^\dagger_\kappa\,dw_\kappa
 \label{ngxxviii}
\ee
are {\it gauge covariant},
\be
 B_\kappa(z)\rightarrow e^{i\theta(x)}B_\kappa e^{i\theta(x)}\ ,\quad
 B_\kappa(z)^*\rightarrow e^{-i\theta(x)}B_\kappa^*e^{-i\theta(x)}
 \label{ngxxix}
\ee
under $z\rightarrow z e^{i\theta}$.

We can account for $U(x)$ by considering
\beqa
{\cal V}_\kappa=Uv_\kappa\ &,&\quad {\cal A}_\kappa={\cal
V}_\kappa^\dagger\,d{\cal V}_\kappa\ ,\quad
 {\cal D}_\kappa=d+ {\cal A}_\kappa\ ,\quad{\cal D}_\kappa^2=d{\cal
A}_\kappa\nn\\
 &\,&{\cal W}_\kappa=(\tau_2U^*\tau_2)w_\kappa\ ,\quad
 {\cal B}_\kappa={\cal W}_\kappa^\dagger\,d{\cal V}_\kappa \ .
 \label{ngxxx}
 \eeqa
${\cal A}_\kappa$ is still a connection, and the properties (\ref{ngxxix})
are not affected by $U$. ${\cal P}_\kappa$ is the support projector 
of ${\cal V}_\kappa^\dagger$, and 
\be
 {\cal W}_\kappa{\cal W}_\kappa^\dagger=\BI-{\cal P}_\kappa\ ,\quad
 (\BI-{\cal P}_\kappa){\cal V}_\kappa=0 \ .
 \label{ngxxxiv}
\ee

Gauge invariant quantities being functions on $S^2$, we can contemplate
a formulation of the $\CP^1$-model as a gauge theory. Let ${\cal J}_i$
be the lift of $L_i$ to angular momentum generators appropriate for 
functions of $z$,
\be
 (e^{i\theta_i{\cal J}_i}f)(z)=f(e^{-i\theta_i\tau_i/2}z)\; ,
 \label{ngxxxi}
\ee
and let
\be
 {\cal B}_{\kappa,i}={\cal W}_\kappa^\dagger\,{\cal J}_i {\cal V}_\kappa\; .
 \label{ngxxxii}
\ee
Now, ${\cal W}_\kappa{\cal B}_{\kappa,i}{\cal V}_\kappa^\dagger$ is gauge 
invariant, and should have an expression in terms of ${\cal P}_\kappa$. 
Indeed it is, in view of (\ref{ngxxxiv}),
\be
{\cal W}_\kappa{\cal B}_{\kappa,i}{\cal V}_\kappa^\dagger=
{\cal W}_\kappa{\cal W}_\kappa^\dagger ({\cal J}_i{\cal V}_\kappa){\cal
V}_\kappa^\dagger=
(\BI-{\cal P}_\kappa){\cal J}_i({\cal V}_\kappa{\cal V}_\kappa^\dagger)=
(\BI-{\cal P}_\kappa)({\cal L}_i{\cal P}_\kappa)=({\cal L}_i{\cal
P}_\kappa){\cal P}_\kappa\; .
\label{ngxxxv}
\ee
Therefore we can write the action (\ref{ngxv}, \ref{ngxvi}) in terms of 
the ${\cal B}_{\kappa,i}$:
\beqa
S_\kappa&=&-
2c\int_{S^2}d\Omega\tr\,{\cal P}_\kappa({\cal L}_i{\cal P}_\kappa)
({\cal L}_i{\cal P}_\kappa)=
2c\int_{S^2}d\Omega\tr\,
(({\cal L}_i{\cal P}_\kappa){\cal P}_\kappa )^\dagger
(({\cal L}_i{\cal P}_\kappa){\cal P}_\kappa)=\nn\\
&=&
2c\int_{S^2}d\Omega\tr({\cal W}_\kappa{\cal B}_{\kappa,i}
{\cal V}_\kappa^\dagger)^\dagger
({\cal W}_\kappa{\cal B}_{\kappa,i}{\cal V}_\kappa^\dagger)=
2c\int_{S^2}d\Omega\;{\cal B}_{\kappa,i}^*{\cal B}_{\kappa,i}\ .
\label{ngxxxvi}
\eeqa

It is instructive also to write the gauge invariant $(d{\cal A}_\kappa)$
in terms of ${\cal P}_\kappa$ and relate its integral to the winding number
(\ref{ngvi}). The matrix of forms
\be
 {\cal V}_\kappa(d+{\cal A}_\kappa){\cal V}_\kappa^\dagger
 \label{ngxxxvii}
\ee
is gauge invariant. Here
\be
d{\cal V}_\kappa^\dagger=(d{\cal V}_\kappa^\dagger)+
{\cal V}_\kappa^\dagger\, d 
\nn\ee
where $d$ in the first term differentiates only ${\cal V}_\kappa^\dagger$.
Now
\be
{\cal V}_\kappa(d+{\cal V}_\kappa^\dagger(d{\cal V}_\kappa))
{\cal V}_\kappa^\dagger
\nn\ee
and
\be
{\cal P}_\kappa\,d{\cal P}_\kappa={\cal V}_\kappa{\cal V}_\kappa^\dagger\,d\,
({\cal V}_\kappa{\cal V}_\kappa^\dagger)=
{\cal V}_\kappa{\cal V}_\kappa^\dagger(d{\cal V}_\kappa)
{\cal V}_\kappa^\dagger+{\cal V}_\kappa(d{\cal V}_\kappa^\dagger)
+{\cal V}_\kappa{\cal V}_\kappa^\dagger\, d
\label{ngxxxviib}
\ee
are equal. Hence, squaring
\be
{\cal V}_\kappa(d+{\cal A}_\kappa)^2{\cal V}_\kappa^\dagger=
{\cal V}_\kappa\,(d{\cal A}_\kappa){\cal V}_\kappa^\dagger=
{\cal P}_\kappa\,(d{\cal P}_\kappa) \,(d{\cal P}_\kappa)
\label{ngxxxviic}
\ee
on using $d^2=0$, eq.(\ref{ngxxxviib}) and 
${\cal P}_\kappa(d{\cal P}_\kappa){\cal P}_\kappa=0$ . Thus
\be 
\int_{S^2}(d{\cal A}_\kappa)=\int_{S^2}\tr\, {\cal V}_\kappa(d{\cal A}_\kappa)
{\cal V}_\kappa^\dagger=\int_{S^2}\tr\,{\cal P}_\kappa\,(d{\cal P}_\kappa)
\,(d{\cal P}_\kappa)\ .
\label{ngxxxviii}
\ee 
We can integrate the LHS. For this we write (taking a section
of the bundle $U(1)\rightarrow S^3\rightarrow S^2$ over
$S^2 \backslash \{ \hbox{north pole} (0,0,1)\})$,
\be 
z(x)=e^{-i\tau_3\varphi/2}e^{-i\tau_2\theta/2}e^{-i\tau_3\varphi/2}
\left(\begin{matrix} 1\\0 \end{matrix}\right)=
\left(\begin{matrix} e^{-i\varphi}\cos\frac{\theta}{2}\\
\sin\frac{\theta}{2} \end{matrix}\right)\;.
\label{ngxxxviiia}
\ee
Taking into account the fact that $U(\vec x)$ is independent of 
$\varphi$ at $\theta=0$, we get
\be
 \int_{S^2}(d{\cal A}_\kappa)=-\int e^{i\kappa\varphi}\, d e^{-i\kappa\varphi}
 =2\pi i \kappa \;.
 \label{ngxxxix}
\ee
This and eq.(\ref{ngxxxviii}) reproduce eq.(\ref{ngvi}).

The Belavin-Polyakov bound \cite{belavin} for $S_\kappa$ can now be got 
from the inequality
\be
\tr {\cal C}_{\kappa,i}^\dagger{\cal C}_{\kappa,i}\ge 0\ ,\quad
 {\cal C}_{\kappa,i}=
{\cal W}_\kappa{\cal B}_{\kappa,i}{\cal V}_\kappa^\dagger\pm
{\cal W}_\kappa(\epsilon_{ijl}x_j{\cal B}_{\kappa,l}){\cal V}_\kappa^\dagger\;.
\label{ngxl}
\ee

\subsection{Connection to an earlier paper.}

In a previous paper \cite{baez}, for $\kappa>0$, the fuzzy $\sigma$-model 
was based on the continuum projector
\be
 P^{(\kappa)}(x)=P_1(x)\otimes...\otimes P_1(x) =\prod_{i=1}^\kappa
 \,\half(1+\tau^{(i)}\cdot x)
 \label{ngxli}
\ee
and its unitary transform
\be
{\cal P}^{(\kappa)}(x)=U^{(\kappa)}(x) P^{(\kappa)}(x)U^{(\kappa)}(x)^{-1}
\ ,\quad
 U^{(\kappa)}(x)=U(x)\otimes...\otimes U(x)\quad (\kappa \hbox{ factors}).
\label{ngxlii}
\ee
At each $x$, the stability group of $P^{(\kappa)}(x)$ is $U(1)$ with
generator $\half\sum_{i=1}^\kappa\tau^{(i)}\cdot x$, and we get a sphere
$S^2$ as $U(x)$ is varied. Thus $U^{(\kappa)}(x)$ gives a section of a
sphere bundle over a sphere, leading us to identify $ {\cal P}^{(\kappa)}$
with a $\CP^1$-field. Furthermore, the R.H.S. of eq.(\ref{ngxxxviii}) 
(with ${\cal P}^{(\kappa)}$ replacing ${\cal P}_\kappa$) gives $\kappa$ as 
the invariant associated with ${\cal P}^{(\kappa)}$, suggesting a 
correspondence between $\kappa$ and winding number. 

We can write
\be  
 {\cal P}^{(\kappa)}= {\cal V}^{(\kappa)} {\cal V}^{(\kappa)\dagger}\ ,\quad
 {\cal V}^{(\kappa)}={\cal V}_1\otimes...\otimes{\cal V}_1
 \quad \kappa \hbox{ factors}),
 \label{ngxliii}
\ee
its connection $ {\cal A}^{(\kappa)}$ and an action as previously.
A computation similar to the one leading to eq.(\ref{ngxxxviii}) shows that
\be
-\frac{i}{2\pi}\int\, d{\cal A}^{(\kappa)}=\kappa\; .
 \label{ngxlix}
\ee
So $\kappa$ is the Chern invariant of the projective module associated with
${\cal P}^{(\kappa)}$. 

For $\kappa<0$, we must change $x$ to $-x$ in (\ref{ngxli}),
and accordingly change other expressions.

But we missed the fact that $\kappa$ cannot be identified with the winding
number of the map $x\rightarrow{\cal P}_\kappa(x)$. To see this, say for 
$\kappa>0$, we show that there is a winding number $\kappa$ map  from  
${\cal P}^{(\kappa)}$ to ${\cal P}_\kappa(x)$. As that is also the winding
number of the map $x\rightarrow{\cal P}_\kappa(x)$, the map 
$x\rightarrow{\cal P}^{(\kappa)}(x)$ must have winding number 1.

The map ${\cal P}^{(\kappa)}\rightarrow{\cal P}_\kappa(x)$ is induced from the
map
\be  
 {\cal V}^{(\kappa)}\ \rightarrow\  {\cal V}_\kappa=
\left(\begin{matrix} {\cal V}^{(\kappa)}_{11...1}\\
{\cal V}^{(\kappa)}_{22...2}\end{matrix}\right)
 \label{ngl}
\ee
and their expressions in terms of ${\cal V}^{(\kappa)}$ and ${\cal V}_\kappa$.
In (\ref{ngl}) all the points \\
${\cal V}^{(\kappa)}(z_1e^{2\pi i\,j/\kappa},z_2e^{2\pi i\,l/\kappa})$,
$j,l\in\{0,1,...,\kappa-1\}$, have the same image,
but in the passage to ${\cal P}^{(\kappa)}$ and ${\cal P}_\kappa$ the overall
phase
of $z$ is immaterial. However, the projectors for
${\cal V}^{(\kappa)}(z_1e^{2\pi ij/\kappa},z_2)$ and 
${\cal V}_\kappa^\dagger(z_1,z_2e^{2\pi ij/\kappa})$ are distinct and map to the
same
${\cal P}_\kappa$, giving winding number $\kappa$.

We have not understood the relation between the models based on  
${\cal P}^{(\kappa)}$ and ${\cal P}_\kappa$.

\section{Fuzzy $\CP^1$-models}
\setcounter{equation}{0}

The advantage of the preceding formulation using $\{z_\alpha\}$ is that the 
passage to fuzzy models is relatively transparent. Thus let
$\xi=(\xi_1,\xi_2)\in{\mathbb C}^2\backslash\{0\}$. 
We can then identify $z$ and $x$ as
 \be 
 z=\frac{\xi}{|\xi|}\ ,\quad |\xi|=\sqrt{|\xi_1|^2+|\xi_2|^2}\ ,\qquad
 x_i=z^\dagger\tau_i z\ .
 \label{ngli}
\ee

Quantization of the $\xi$'s and $\xi^*$'s consists in replacing 
$\xi_\alpha$ by annihilation operators $a_\alpha$ and $\xi_\alpha^*$ 
by $a_\alpha^\dagger$.
$|\xi|$ is then the square root of the number operator:
\beqa
\hat N&=&\hat N_1+\hat N_2\ ,\quad 
\hat N_1=a^\dagger_1 a_1\ ,\ N_2=a^\dagger_2 a_2\ ,\nn\\ 
\hat z^\dagger_\alpha&=&\frac{1}{\sqrt{\hat N}}
a^\dagger_\alpha=a^\dagger_\alpha\frac{1}{\sqrt{\hat N +1}}\ ,\quad
\hat z_\alpha=\frac{1}{\sqrt{\hat N +1}}a_\alpha=
a_\alpha\frac{1}{\sqrt{\hat N}}\;,\nn\\
\hat x_i&=&\frac{1}{\sqrt{\hat N}}a^\dagger\tau_i a\ .
 \label{nglii}
 \eeqa  
(We have used hats on some symbols to distinguish them as fuzzy operators).

We will apply these operators only on the subspace of the Fock space
with eigenvalue $n$ of $\hat N$, $\ge 1$, where $\frac{1}{\sqrt{\hat N}}$ is
well defined. This restriction is natural and reflects the fact that $\xi$
cannot be zero.

\subsection{ The fuzzy projectors for $\kappa>0$\; .}.
 On referring to (\ref{ngix}), we see that if $\kappa>0$, for the
quantized versions $\hat v_\kappa,\;\hat v^\dagger_\kappa$ of 
$v_\kappa,\; v^*_\kappa$, we have
\beqa
\hat v_\kappa&=&\left[ \begin{matrix}a_1^\kappa\\a_2^\kappa\end{matrix}\right]
\frac{1}{\sqrt{\hat Z_\kappa}}\ ,\qquad
\hat v_\kappa^\dagger=\frac{1}{\sqrt{\hat Z_\kappa}}
\left[ \begin{matrix}
(a_1^\dagger)^\kappa&(a_2^\dagger)^\kappa\end{matrix}\right]
\ ,\qquad \hat v_\kappa^\dagger\hat v_\kappa=\BI\ ,\nn\\
\hat Z_\kappa&=&\hat Z_\kappa^{(1)}+\hat Z_\kappa^{(2)}\quad ,\quad
\hat Z_\kappa^{(\alpha)}=
\hat N_\alpha(\hat N_\alpha-1)...(\hat N_\alpha-\kappa+1)\ ..
\label{ngliii}
\eeqa

The fuzzy analogue of $U$ is a $2\times 2$ unitary matrix $\hat U$
whose entries $\hat U_{ij}$ are polynomials in $a^\dagger_a a_b$. 
As for $\hat{\cal V}_\kappa$, the quantized version of ${\cal V}_\kappa$, it is
just
\be
 \hat{\cal V}_\kappa=\hat U\,\hat v_\kappa
 \label{ngliv}
\ee
and fulfills
\be
 \hat{\cal V}_\kappa^\dagger\,\hat{\cal V}_\kappa=\BI\; ,
 \label{nglv}
\ee
$ \hat{\cal V}_\kappa^\dagger$ being the quantized version of 
${\cal V}_\kappa^\dagger$.
We thus have the fuzzy projectors
\be
\hat P_\kappa=\hat v_\kappa\hat\, v_\kappa^\dagger\ ,\qquad
 \hat {\cal P}_\kappa=\hat{\cal V}_\kappa\,\hat{\cal V}_\kappa^\dagger\;.
 \label{nglvi}
\ee 

Unlike $\hat v_\kappa,\hat{\cal V}_\kappa $ and their adjoints, 
$\hat P_\kappa$ and $\hat {\cal P}_\kappa$ commute with the number operator
$\hat N$. So we can formulate a finite-dimensional matrix model
for these projectors as follows. 
Let ${\cal F}_n$ be the subspace of the Fock 
space where $\hat N=n$. It is of dimension $n+1$, and carries the $SU(2)$
representation with angular momentum $n/2$, the $SU(2)$ generators being
\be
 L_i=\frac{1}{2}a^\dagger\tau_i a \ .
 \label{nglvii}
\ee
Its standard orthonormal basis is $|\frac{n}{2},m>\;,\ m=-\frac{n}{2},
-\frac{n}{2}+1,...,\frac{n}{2}$.
Now consider ${\cal F}_n\otimes_{\mathbb C}{\mathbb C}^2:={\cal F}_n^{(2)}$,
with elements $f=(f_1,f_2),\;f_a\in{\cal F}_n$. Then 
$\hat P_\kappa,\; \hat {\cal P}_\kappa$ act on ${\cal F}_n^{2(2)}$ in the
natural way.
For example
\be
 f\rightarrow\hat {\cal P}_\kappa f, \quad
(\hat{\cal P}_\kappa f)_a=(\hat{\cal P}_\kappa)_{ab}f_b=
 (\hat{\cal V}_{\kappa,a}\hat{\cal V}_{\kappa,b}^\dagger)f_b \ .
 \label{nglviii}
\ee

We now can write explicit matrices for $\hat P_\kappa$ and 
$\hat {\cal P}_\kappa$. We have:
\beqa
\hat P_\kappa&=&\left( \begin{matrix}
a_1^\kappa\frac{1}{\hat Z_\kappa}a_1^{\dagger\,\kappa}
&a_1^\kappa\frac{1}{\hat Z_\kappa}a_2^{\dagger\,\kappa}\\
a_2^\kappa\frac{1}{\hat Z_\kappa}a_1^{\dagger\,\kappa}&
a_2^\kappa\frac{1}{\hat Z_\kappa}a_2^{\dagger\,\kappa}\end{matrix}\right)
\ ,\label{nglix}\\
a_1^\kappa\frac{1}{\hat Z_\kappa}&=&\frac{1}{(\hat N_1+\kappa)...(\hat N_1+1)
+\hat Z^{(2)}_\kappa}a_1^\kappa\ ,\quad
a_1^\kappa a_1^{\dagger\,\kappa}=(\hat N_1+\kappa)...(\hat N_1+1)\ ,\nn
\eeqa 
from which its matrix $\hat P_\kappa(n)$ for $\hat N=n$ can be obtained.

The matrix $ \hat {\cal P}_\kappa$ is the unitary transform 
$\hat U\hat P_\kappa(n)\hat U^\dagger$ where $\hat U$ is a 
$2\times 2$ matrix and 
$\hat U_{ab}$ is itself an $(n+1)\times (n+1)$ matrix.
As for the fuzzy analogue of ${\cal L}_i$, we define it by
\be
{\cal L}_i\hat {\cal P}_\kappa=[L_i,\hat {\cal P}_\kappa]\ .
\label{nglx}
\ee

The fuzzy action 
\be
 S_{F,\kappa}(n)=\frac{c}{2(n+1)}\tr_{\hat N=n}\,({\cal L}_i
\hat {\cal P}_\kappa)^\dagger
 ({\cal L}_i\hat {\cal P}_\kappa)\ ,\quad c=\hbox{constant}\ ,
 \label{nglxi}
\ee
follows, the trace being over the space ${\cal F}_n^{(2)}$.

\subsection{ The Fuzzy Projector for $\kappa<0$\;.}

For $\kappa<0$, following an early indication, we must exchange the roles of
$a_a$ and $a^\dagger_a$.

\subsection{ Fuzzy Winding Number\;.}

In the literature \cite{grosse}, there are suggestions on how to 
extend (\ref{ngvi}) to
the fuzzy case. They do not lead to an integer value for this number
except in the limit $n\rightarrow\infty$.

There is also an approach to topological invariants using Dirac 
operator and 
cyclic cohomology. Elsewhere this approach was applied to the fuzzy case
\cite{baez,balsachin} and gave integer values, and even a fuzzy
analogue of the Belavin-Polyakov bound.  However they were not for 
the action $S_{F,\kappa}$, 
but for an action which approaches it as $n\rightarrow\infty$.
Below, in section 5.4, we present an alternative
approach to this bound which works for $S_{F,\kappa}$. It looks like 
 (\ref{ngxvib}), except that $\kappa$ becomes an integer only in the limit
$n\rightarrow\infty$.

There is also a very simple way to associate an integer to 
$\hat{\cal V}_\kappa$ \cite{peter, sachin}. It
is equivalent to the Dirac operator approach. We can assume that the 
domain of 
$\hat{\cal V}_\kappa$ are vectors with a fixed value $n$ of $\hat N$. 
Then after
applying $\hat{\cal V}_\kappa$, $n$ becomes $n-\kappa$ if $\kappa>0$ and 
$n+|\kappa|$ is $\kappa<0$.
Thus $\kappa$ is just the difference in the value of $\hat N$, or equivalently
twice the difference in the value of the angular momentum, between its 
domain and its range.

We conclude this section by deriving the bound for $S_{F,\kappa}(n)$.

\subsection{The Fuzzy Bound.}

A proper generalization of the Belavin-Polyakov bound to its fuzzy version
involves a slightly more elaborate approach. This is because the 
straightforward fuzzification
of $\vec\sigma\cdot\vec x$ and $\vec\tau\cdot\vec n^{(\kappa)}$
and their corresponding projectors do not commute, and the product of such
fuzzy projectors is not a projector. {\it We use this elaborated approach
only in this section.} It is not needed elsewhere. In any case, what is 
there in other sections is trivially adapted to this formalism.

The approach taken here is not new. It is essential, and has been widely used,
for example for the study of the fuzzy Dirac operator \cite{bal1}.

The operators $a^\dagger_\alpha a_\beta$ acting on the vector space with
$\hat N=n$ generate the algebra $Mat(n+1)$ of $(n+1)\times(n+1)$ matrices.
The extra structure comes from regarding them not as observables, but
as a Hilbert space of matrices $m,\ m',...$ with scalar product
$(m',m)=\frac{1}{n+1}\tr_{{\mathbb C}^{n+1}}{m'}^\dagger\;m$, with the 
observables acting thereon.

To each $\alpha\in Mat(n+1)$, we can associate two linear operators
 $\alpha^{L, R}$ on  $Mat(n+1)$  according to
\be
\alpha^Lm=\alpha m\ ,\quad \alpha^R m=m\alpha\ ,\quad m\in\;Mat(n+1)\ .
\label{ngci}
\ee
$\alpha^L-\alpha^R$ has a smooth commutative limit for operators
of interest. It actually vanishes, and $\alpha^{L, R}\to 0$ if $\alpha$
remains bounded during this limit.

Consider the angular momentum operators $L_i\in Mat(n+1)$. The associated
`left' and `right' angular momenta $L_i^{L,R}$ fulfill
\be
(L_i^L)^2=(L^R_i)^2=\frac{n}{2}(\frac{n}{2}+1)\ .
\label{ngcii}
\ee

We now regard $a_\alpha,\ a_\alpha^\dagger$ of section 5 as left operators
$a_\alpha^L$ and $ a_\alpha^{\dagger L}$. $\hat P_\kappa^L$ thus becomes 
a $2\times 2$ matrix with each entry being a left multiplication
operator. It is the linear operator $\hat{\cal P}^L_\kappa$
on $Mat(n+1)\otimes{\mathbb C}^2$.
We tensor this vector space with another ${\mathbb C}^2$ as before to get
${\cal H}=Mat(n+1)\otimes{\mathbb C}^2\otimes{\mathbb C}^2$, with
$\sigma_i$ acting on the last ${\mathbb C}^2$, and 
 $\sigma\cdot{\cal L}\hat{\cal P}^L_\kappa$ denoting the operator
 $\sigma_i({\cal L}_i\hat{\cal P}_\kappa)^L$.

We can repeat the previous steps if there are fuzzy analogues $\gamma$
and $\Gamma$ of continuum `world volume' and `target space' chiralities
$\vec\sigma\cdot\vec x$ and $\vec\tau\cdot\vec n^{(\kappa)}$ which
mutually commute. Then $\frac{1}{2}(1\pm\gamma)$, $\frac{1}{2}(1\pm\Gamma)$
are commuting projectors and the expressions derived at the end of Section 3
 generalize, as we shall see.

There is such 
a $\gamma$, due to Watamuras\cite{watamura}, and  discussed further
by \cite{baez}. Following \cite{baez}, we take
\be
\gamma\equiv\gamma^L=\frac{2\sigma\cdot L^L+1}{n+1}\ .
\label{ngciv}
\ee
The index $L$ has been put to emphasize its left action on $Mat(n+1)$.

As for $\Gamma$, we can do the following. $\hat{\cal P}_\kappa$ acts on the
left on $Mat(n+1)$, let us call it $\hat{\cal P}_\kappa^L$. It has a
$\hat {\cal P}_\kappa^R$ acting on the right and an associated
\be
\Gamma\equiv\Gamma^R_\kappa=2\hat{\cal P}_\kappa^R-1\quad ,
\quad(\Gamma^R_\kappa)^2
=1\ .
\label{ngcv}
\ee
As it acts on the right and involves $\tau$'s while $\gamma$ acts on the left
and involves $\sigma$'s,
\be
\gamma^L\Gamma^R_\kappa=\Gamma^R_\kappa\gamma^L\ .
\label{ngcvi}
\ee

The bound for (\ref{nglxi}) now follows from
\be
\tr_{\cal H}\left(
\frac{1+\epsilon_1\gamma^L}{2}\frac{1+\epsilon_2\Gamma^R_\kappa}{2}
\sigma\cdot{\cal L}\hat{\cal P}_\kappa^L
\right)^\dagger
\left(
\frac{1+\epsilon_1\gamma^L}{2}\frac{1+\epsilon_2\Gamma^R_\kappa}{2}
\sigma\cdot{\cal L}\hat{\cal P}_\kappa^L
\right)\ge 0
\label{ngcvii}
\ee
($\epsilon_1,\epsilon_2=\pm 1$), and reads
\beqa
S_{F,\kappa}&=&
\frac{c}{4(n+1)}\tr_{\cal H}(\sigma\cdot{\cal L}\hat{\cal P}_\kappa^L)^\dagger
(\sigma\cdot{\cal L}\hat{\cal P}_\kappa^L)\nn\\
&\ge&\frac{c}{4(n+1)}\tr_{\cal H}\left((\epsilon_1\gamma^L+
\epsilon_2\Gamma^R_\kappa)(\sigma\cdot{\cal L}\hat{\cal P}_\kappa^L)
(\sigma\cdot{\cal L}\hat{\cal P}_\kappa^L)\right)\nn\\
&&+\,\frac{c}{4(n+1)}\tr_{\cal H}\left(\epsilon_1\epsilon_2\gamma^L\Gamma^R
(\sigma\cdot{\cal L}\hat{\cal P}_\kappa^L)
(\sigma\cdot{\cal L}\hat{\cal P}_\kappa^L)
\right)
\label{ngcviii}
\eeqa
The analogue of the first term on the R.H.S. is zero in the continuum,
being absent in (\ref{ngxvib}), but not so now. As $n\to\infty$,
 (\ref{ngcviii}) reproduces (\ref{ngxvib}) to leading order $n$, 
but has corrections 
which vanish in the large $n$ limit.

A minor clarification: if $\tau$'s are substituted by $\sigma$'s in
$2\hat{\cal P}^L_1-1$, then it is $\gamma_L$. The different projectors
are thus being constructed using the same principles.

\section{$\CP^N$-Models}
\setcounter{equation}{0}

We need a generalization of the Bott projectors to adapt the previous
approach to all $\CP^N$.

Fortunately this can be easily done. The space $\CP^N$
is the space of $(N+1)\times(N+1)$ {\it rank 1} projectors.  
The important point is the rank.  So we can write
\be
 \CP^N=<U^{(N+1)}P_0U^{(N+1)\dagger}:\  P_0=
\hbox{diag.}\underbrace{(0,....,0,1)}_{N+1 \;entries}\ ,
 \ U^{(N+1)}\in U(N+1)>\;.
 \label{nglxii}
\ee

As before, let $z=(z_1,z_2),\ |z_1|^2+|z_2|^2=1$, and 
$x_i=z^\dagger\tau_iz$.
Then we define
\be
v_\kappa^{(N)}(z)=\left(
\begin{matrix}z_1^\kappa\\z_2^\kappa\\0\\.\\.\\0\end{matrix}
 \right)\frac{1}{\sqrt{Z_\kappa}}\;,\ \kappa>0\,;\quad v_\kappa^{(N)}(z)
 =\left(\begin{matrix}z_1^{*\kappa}\\z_2^{*\kappa}\\0\\.\\.\\0\end{matrix}
 \right)\frac{1}{\sqrt{Z_\kappa}}\;,\ \kappa<0\;. 
 \label{nglxiii}
\ee
Since
\beqa
 v_\kappa^{(N)}(z)^\dagger v_\kappa^{(N)}(z)&=&1\,,\nn\\
 P^{(N)}_\kappa(x)&=&v_\kappa^{(N)}(z)v_\kappa^{(N)}(z)^\dagger\,\in\,\CP^N\; .
 \label{nglxiv}
\eeqa

We can now easily generalize the previous discussion, using 
$P^{(N)}_\kappa$ for $P_\kappa$ and $U^{(N+1)}$ 
for $U$, and subsequently quantizing $z_\alpha,\,z_\alpha^*$. 
In that way we get fuzzy $\CP^N$-models.

$\CP^N$ models can be generalized by replacing the target space
 by a general Grassmannian or a flag manifold. They
can also be elegantly formulated as gauge theories \cite{baletal}. 
But we are able to formulate only a limited class of such manifolds
in such a way that they can be made fuzzy.
The natural idea would be to look for several vectors
\be
 v_{k_i}^{(N)(i)}(z)\ ,\quad i=1,..,N
 \label{nglxv}
\ee 
in $(N+1)$ dimensions which are normalized and orthogonal,
\be
 v_{k_i}^{(N)(i)\dagger}(z) v_{k_j}^{(N)(j)}(z) =\delta_{ij}
 \label{nglxvi}
\ee
and have the equivariance property
\be
  v_{k_i}^{(N)(i)}(ze^{i\theta})= v_{k_i}^{(N)(i)}(z)e^{i\,k_i\theta}
 \label{nglxvii}
\ee
The orbit of the projector $\sum_{i=1}^M v_{k_i}^{(N)(i)}(z) 
v_{k_i}^{(N)(i)\dagger}(z)$
under $U^{(N+1)}$ will then be a Grassmannian for each $M\le N$, while the 
orbit of $\sum_i \lambda_iv_{k_i}^{(N)(i)}(z) v_{k_i}^{(N)(i)\dagger}(z)$ 
with possibly unequal
$\lambda_i$ under $U^{(N+1)}$ will be a flag manifold.

But we can find such $ v_{k_i}^{(N)(i)}$ only for 
$i=1,2,...,M\le\frac{N+1}{2}$.

For instance in an $(N+1)=2L$-dimensional vector space, for integer $L$, 
we can form the vectors
\be
v_{k_1}^{(N)(1)}(z)=\left(\begin{matrix}z_1^{k_1}\\z_2^{k_1}\\0\\ \cdot\\0
\end{matrix}\right)\frac{1}{\sqrt{Z_{k_1}}}\ ,\ 
v_{k_2}^{(N)(2)}(z)=
\left(\begin{matrix}0\\0\\z_1^{k_2}\\z_2^{k_2}\\0\\ \cdot\\0
\end{matrix}\right)\frac{1}{\sqrt{Z_{k_2}}}\ ,\ ...\ ,\ 
v_{k_L}^{(N)(L)}(z)=\left(\begin{matrix}0\\ \cdot\\0\\z_1^{k_L}\\z_2^{k_L}
\end{matrix}\right)\frac{1}{\sqrt{Z_{k_L}}}
\label{nglxviii}
\ee
for $k_i>0$. For those $k_i$ which are negative, we replace 
$v^{(N)(i)}_{k_i}(z)$ here by $v^{(N)(i)}_{|k_i|}(z)^*$:
\be
v^{(N)(i)}_{k_i}(z)=v^{(N)(i)}_{|k_i|}(z)^*\ ,\ k_i<0\ .
\label{nglxix}
\ee
These $v^{(N)(i)}_{k_i}$ are orthonormal for all $z$ with 
$\sum_\alpha |z_\alpha|^2=1$, so that we can handle Grassmannians and
flag manifolds involving projectors up to rank $L$.

If $N$ instead is $2L$, we can write
\be
v_{k_1}^{(N)(1)}(z)=\left(\begin{matrix}z_1^{k_1}\\z_2^{k_1}\\0\\ \cdot\\0
\end{matrix}\right)\frac{1}{\sqrt{Z_{k_1}}}\ ,\ 
v_{k_2}^{(N)(2)}(z)=
\left(\begin{matrix}0\\0\\z_1^{k_2}\\z_2^{k_2}\\0\\ \cdot\\0
\end{matrix}\right)\frac{1}{\sqrt{Z_{k_2}}}\ ,\ ...\ ,\ 
v_{k_L}^{(N)(L)}(z)=\left(\begin{matrix}0\\ \cdot\\0\\z_1^{k_L}\\z_2^{k_L}\\
0\end{matrix}\right)\frac{1}{\sqrt{Z_{k_L}}}
\label{nglxviiib}
\ee
for $k_i>0$, and use (\ref{nglxix}) for $k_i<0$. 

But we can find no
vector $v_{k_{L+1}}^{(N)(L+1)}(z)$ fulfilling
\be
v_{k_i}^{(N)(i)}(z)^\dagger v_{k_{L+1}}^{(N)(L+1)}(z)=\delta_{i,L+1},\ 
i=1,2,..,L+1 \ ,\quad
v_{k_{L+1}}^{(N)(L+1)}(ze^{i\theta})=v_{k_{L+1}}^{(N)(L+1)}(z)
e^{ik_{L+1}\theta}\ . 
\label{nglxixa}
\ee

The quantization or fuzzification of these models can be done as before.

But lacking suitable $v^{(i)}_{k_i}$ for $i>L$, the method fails if the 
target flag manifold involves projectors of rank $>\frac{N+1}{2}$.

Note that we cannot consider vectors like
\be
v'(z)=\left(\begin{matrix}0\\ \cdot\\0\\z_i^{k}\\0\\ \cdot\\0
\end{matrix}\right)\frac{1}{|z_i|^k}\ ,\quad k>0\ ,\ i=1\;or\;2
\label{nglxxx}
\ee
and $v'(z)^*$. That is because $z_i$ can vanish compatibly with the 
constraint
$|z_1|^2+|z_2|^2=1$, and $v'(z),\, v'(z)^*$ are ill-defined when $z_i=0$.
 
As mentioned before, the flag manifolds are coset spaces
${\cal M}=SU(K)/S(U(k_1)\otimes U(k_2)\otimes..\otimes U(k_\sigma)),
\ \sum k_i=K$. Since $\pi_2({\cal M})=
\underbrace{{\mathbb Z}\oplus...\oplus{\mathbb Z}}_{\sigma\; terms}$,
a soliton on ${\cal M}$ is now characterized by $\sigma$ winding numbers,
with each number allowed to take either sign. The two possible signs
for $k_i$ in $v^{(i)}_{k_i}$ reflect this freedom.

\vspace{0.5cm}
{\bf Acknowledgements}

We are part of a collaboration working on fuzzy physics which 
involves several physicists. The work reported here has benefited 
greatly by discussions with the members of this collaboration. It was
supported by D.O.E. and N.S.F., U.S.A. under contract numbers 
DE - FG02 - 85ER40231 and INT - 9908763, and by I.N.F.N., Italy. 

\bigskip\bigskip\bigskip\bigskip
\bibliographystyle{unsrt}

\end{document}